\documentclass[10pt,preprint]{aastex}
\usepackage{emulateapj5}
\usepackage{amsmath,natbib}

\begin{document}
%------------------------------------------------------------------------------

\title{Iron-Line Emission as a Probe of Bardeen-Petterson Accretion Disks}

\author{P. Chris Fragile}
\affil{Physics Department, University of California, Santa Barbara,
CA 93106}

\and

\author{Warner A. Miller, Eric Vandernoot}
\affil{Physics Department, Florida Atlantic University, Boca Raton,
FL 33431}

%-----------------------------------------------------------------------------

\begin{abstract}
In this work we show that Bardeen-Petterson accretion disks can 
exhibit unique, detectable features in relativistically broadened 
emission line profiles. Some of the unique characteristics include 
inverted line profiles with sharper red horns and softer blue horns 
and even profiles with more than 2 horns from a single rest-frame line.
We demonstrate these points by constructing 
a series of synthetic line profiles using 
simple two-component disk models. We find that the resultant profiles 
are very sensitive to the two key parameters one would
like to constrain, namely the Bardeen-Petterson transition radius
$r_{BP}$ and the relative tilt $\beta$ between the two disk
components over a range of likely values [$10\le r_{BP}/(GM/c^2) \le
40 ; 15^\circ \le \beta \le 45^\circ$]. We use our findings to show
that some of the ``extra'' line features observed in the spectrum of
the Seyfert-I galaxy MCG--6-30-15 may be attributable to a
Bardeen-Petterson disk structure. Similarly, we apply our findings to two
likely Bardeen-Petterson candidate Galactic black holes - GRO
J1655-40 and XTE J1550-564. We provide synthetic line profiles of these
systems using observationally constrained sets of parameters.
Although we do not formally fit the data for any of these systems,
we confirm that our synthetic spectra are consistent with current
observations.
\end{abstract}


\section{Introduction}
\label{sec:introduction} Relativistically broadened emission lines
have been observed in the X-ray spectra of over a dozen active
galactic nuclei (AGN) \citep{mus95,tan95,nan97}. More recently
similar features have also been discovered in Galactic black holes
(GBHs) and black hole candidates \citep{mil02b,mil02c,mil02a}. These
lines, from fluorescent K$\alpha$ emission of iron, provide a unique
diagnostic of the inner regions of accretion flows around black
holes.  Such fluorescent lines are produced when hard X-rays
illuminate regions of optically thick disk material
\citep{geo91,mat91}. Since the line energies are well known (6.4 keV
for cold iron with ionization states below Fe XVII, 6.7 keV for Fe
XXV, and 6.9 keV for Fe XXVI) and the intrinsic widths are very
small, the observed, broadened line profiles can provide direct
information on the Doppler shifts and gravitational redshifts
affecting the line-emitting material \citep{fab89,lao91}. If one
assumes the emitting gas to be in the form of a thin Keplerian disk,
then these line profiles also provide information on the emissivity
of the disk, the radial distribution of the emitting gas, and the
observed inclination angle $i$ of the disk ($i=0^\circ$ is face on)
\citep{lao91,par98,par01}. The profiles of these lines are also
sensitive to features of more realistic disk models such as
turbulent broadening and radial inflow \citep{par98,arm03}.

The dependence of these line profiles on the radius of emission and
inclination of the disk suggest that they may be useful as a
diagnostic of warped or twisted accretion disks \citep{har00}.
%As an example, the
%Sgr gives a best-fit inclination angle of $i=43^\circ \pm 15^\circ$
%for the inner disk \citep{mil02a}, whereas the optically determined
%inclination is $60^\circ \lesssim i \le 71^\circ$ \citep{oro01},
%suggesting a possible warp in the disk.
One such warping mechanism which is active at small radii around
rapidly rotating black holes is caused by relativistic
Lense-Thirring precession. Here the frame dragging of the rotating
black hole causes orbits that are inclined (or tilted) with respect
to the spin axis of the black hole to precess.  For test particles,
the precession frequency is $\Omega_{LT}=2GMa/(c^2r^3)$. For a
tilted disk, the Lense-Thirring precession causes a radially
dependent torque which, by itself, would cause the disk to twist up.
However, in a disk this twisting must compete with whatever angular
momentum transport mechanism (i.e. "viscosity") is driving the
accretion. In such a case, the quasi-steady-state solution is
characterized by a two-component disk with the inner disk aligned
with the symmetry plane of the black hole and the outer disk
essentially retaining its original orientation (as set by the
angular momentum of the gas reservoir). This is the so-called
Bardeen-Petterson effect \citep{bar75}. It is characterized in terms
of a tilt angle $\beta$ between the inner and outer disk components
and a transition radius $r_{BP}$ where the two components connect.

If iron lines are emitted by the disk, then depending on where the
lines are emitted relative to the Bardeen-Petterson transition
radius, the resulting line profile may reveal unique {\em
detectable} spectral features characterizing one or both disk
components. First, if $r_{BP}$ lies within the line emitting region,
part of the line flux will be emitted at one inclination relative to
the observer while the remainder will be emitted at a different
inclination. Since the blue wing ($g_{max}=\nu_{max}/\nu_e$) of a
broadened line profile is sensitive to the inclination of the
emitting region, a tilted, two-component disk could produce multiple
blue horns from a single (rest-frame) line. The red tail
($g_{min}=\nu_{min}/\nu_e$) would also be sensitive to a
Bardeen-Petterson disk since it is responsive to the inner radius of
the emitting region, which would obviously be different for an inner
and outer disk component. The observer would see an integrated line
profile that is in effect a superposition of the line profiles from
the two disk components (minus any reduction in flux due to
shadowing). Thus, as we will show, Bardeen-Petterson disks can exhibit
unique spectral features not seen in flat single-disk models, such
as inverted line profiles with sharper red horns and a softer blue
horns or even profiles with more than 2 horns from a single
(rest-frame) line.

Perhaps the most important possible outcome of this research would
be to directly constrain the scale of the Bardeen-Petterson
transition radius through observations. There is currently great
uncertainty as to where the transition radius is likely to lie, with
estimates varying from $r_{BP} \lesssim 20 r_g$ \citep{nel00} to
$r_{BP} \ge 100 r_g$ \citep{bar75}, where $r_g \equiv GM/c^2$ is the
gravitational radius of the black hole. The expectation is that it
will occur approximately where the rate of twisting by differential
Lense-Thirring precession is balanced by the rate at which twists
are diffused or propagated away by viscosity. The large uncertainty
in $r_{BP}$ then is largely a product of our ignorance of the
details of angular momentum transport in accretion disks,
particularly in tilted disks where the vertical shear component of
viscosity may be crucial. Constraining $r_{BP}$ through observations
would go a long way toward improving our understanding of black hole
accretion disks.

Even if $r_{BP}$ lies outside the line-emitting region, the presence
of a Bardeen-Petterson disk could still be inferred in some cases.
For instance, the tilted outer disk may shadow parts of the inner
line-emitting region.  The observable signatures of such shadowing
was explored in \citet{har00}, although for somewhat different
scales than are considered here. Furthermore, the Bardeen-Petterson
effect provides a natural explanation for why fits of some iron
lines require inclinations that appear inconsistent with other
observations of the same system.  This is seen more often in GBHs
where the inclination of the binary is often well constrained. For
instance, the best-fit models for the iron line in V 4641 Sgr give
an inclination of $i=43^\circ \pm 15^\circ$ \citep{mil02c}, whereas
the optically determined inclination (of the far outer disk) is
$60^\circ \lesssim i \le 71^\circ$ \citep{oro01}. Although these
inclination measurements are marginally consistent with one another,
a true discrepancy of this type could easily be reconciled if one
assumes that the iron line comes from the inner component of a
two-component Bardeen-Petterson disk.

We consider a range of values for the tilt $\beta$ between the
disks, with magnitudes in the range $0^\circ \le | \beta | \le
45^\circ$. For GBHs, $\beta \lesssim 30^\circ$ is consistent with
expectations based upon the formation avenues of compact binaries
\citep[cf.][]{fra01a}.  It is also consistent with the currently
available observational constraints on such systems. For instance,
as noted above, the best-fit models for the iron line in V 4641 Sgr
give an inclination of $i=43^\circ \pm 15^\circ$ \citep{mil02c},
whereas the optically determined inclination 
is $60^\circ \lesssim i \le 71^\circ$ \citep{oro01}, yielding a
possible tilt between the outer disk and the inner
iron-line-emitting region of $\beta > 2^\circ$ (as we discuss below,
the upper limit is uncertain because the orientation of the observer
relative to the line-of-nodes is unknown). Interestingly, the radio
jet observed from this source forms an angle with the line-of-sight
of $\theta < 6^\circ$ \citep{oro01}. If the axis of this jet is
coincident with the spin axis of the black hole or the angular
momentum of the inner accretion flow, as is commonly assumed, then
the tilt could be quite large, $\beta > 54^\circ$. A similar,
although less extreme, discrepancy is observed between the jet and
disk inclinations in GRO J1655-40 \citep{fra01a,oro01}, where the
jet lies almost in the plane of the sky \citep[$\theta = 85^\circ
\pm 2^\circ$,][]{hje95} and the disk has an inclination
$i=70.2^\circ \pm 1.9^\circ$ \citep{gre01}, suggesting a possible
tilt of $\beta \gtrsim 15^\circ$. Another example is found in XTE
J1550-564, for which the apparent superluminal motion of its radio
jets \citep{han01}, constrains $\theta$ to be $<53^\circ$ while the
observed inclination of the outer disk is
$i=72.2^\circ\pm5.2^\circ$, again suggesting a possible tilt of
$\beta \gtrsim 14^\circ$.

The case for Bardeen-Petterson disks around supermassive black holes
is less clear. Nevertheless, many supermassive black holes are
thought to be spinning rapidly \citep{elv02,yu02} and two plausible
scenarios for supplying the necessary tilted accretion flow come to
mind: 1) episodic accretion from tidally disrupted transiting
objects, such as is thought to fuel the accretion of Sgr A*, the
supermassive black hole at the center of the Milky Way; or 2)
accretion following galactic merger events where the supermassive
black hole has not had sufficient time to align with the accreting
matter. Neither case would appear to have a preferred orientation
and thus presents the possibility of a full range of tilts.

The main goal of this paper is to highlight some of the unique
features expected in relativistically broadened iron-lines emitted
by Bardeen-Petterson disks. Although this is a highly idealized
study, we show in \S\ref{sec:models} that the profiles are quite
sensitive to the Bardeen-Petterson disk parameters. Observations of
such lines could therefore be used to set meaningful constraints on
the physics of Bardeen-Petterson disks. In \S\ref{sec:examples}, we
proceed to discuss our results in terms of a few selected GBHs and
AGN. We also illustrate that this technique for studying
Bardeen-Petterson disks through relativistically broadened line
profiles will be most successful when combined with other
observations that can help constrain the system parameters.

\section{Imaging and Line Profiles of Bardeen-Petterson Disks}
\label{sec:models}
%We use geometrized units $G=c=1$ throughout this discussion.

\subsection{Disk Model}
We model the Bardeen-Petterson effect very simplistically as two
separate infinitesimally thin disks: an inner disk extending from $r_{in}$ to
$r_{BP}$ and an outer disk extending from $r_{BP}$ to $r_{out}$. For
most of our models we assume the black hole to be maximally rotating
so that $a/M = Jc/GM^2 =1$. In this case, the formal cut-off for the
disk at the last stable orbit coincides with the horizon of the
black hole, i.e. $r_{ms}= r_{BH} = r_g$. However, we effectively
truncate the disk at $r_{in}=1.05r_{BH}$ by terminating photon
trajectories that penetrate this radius. This is done for numerical 
convenience as the integration times of the photon trajectories 
become infinitely long as they approach the horizon. The choice of 
$a/M=1$ was made somewhat arbitrarily, and although this value 
cannot be reached by an astrophysical black hole, we demonstrate 
below that reducing $a/M$ does not cause any more dramatic change 
to our results than those attributed to other uncertainties in our 
medels.

The outermost disk radius is set
to $r_{out}=100r_g$. The transition radius is varied over the range
$10 \le r_{BP}/r_g \le 40$, consistent with the numerical results of
\citet{nel00}. In such a simple two-component Bardeen-Petterson disk
model, we expect separate red and blue horns to be produced by 
each disk component. This is where the power of this technique
lies. Since the blue horns will be sensitive to the relative tilt of
the two disks and the red horns will be sensitive to the location of
the transition radius, the full line profile gives information
about the two parameters one would most like to constrain.

All of the models in this work assume a twist-free disk (i.e. the
line-of-nodes of the two disks is a straight line).  Based upon
numerical simulations \citep{nel00} this seems consistent with
expectations for a Bardeen-Petterson disk.  We generalize our
results by allowing the observer to be placed anywhere on a sphere
of large radius ($500r_g$). The observer's location is described by
two angles \{$\theta_0$, $\phi_0$\}, in the usual Boyer-Lindquist
spherical coordinates centered on the black hole.  Since the
observer's location is often unconstrained, we consider a range of
choices for $\theta_0$ and $\phi_0$.  However, as we show in \S
\ref{sec:bhbs}, if the inclination of the inner or outer disk
relative to the observer can be constrained, then the observer's
location is no longer arbitrary.

\subsection{Ray Tracing}
To obtain an image of the accretion disk as seen by a distant
observer, we integrate the geodesic equations of motion using a
variant of the numerical ray tracing code described in
\citet{bro97}.  We follow the trajectories of photons from a
pixelated grid of points at some large distance ($r=500r_g$) back to
the accretion disk. The uniform grid may be associated with the
field imaged by a CCD detector for instance. The observed frequency
at each pixel is given by
\begin{equation}
g \equiv \frac{\nu_o}{\nu_e}=\frac{-1}{-u \cdot p} ~,
\end{equation}
where subscripts $o$ and $e$ are observer and emitter, respectively,
$u$ is the 4-velocity of the emitter, and $p$ is the emitted
photon's 4-momentum.  Generally, the emitter 4-velocity is specified
by the disk model.  In this work we simply use the Keplerian
4-velocity in the plane of the disk, ignoring effects such as radial
transport and turbulent motion which have been explored in a general
context elsewhere \citep{par98}. The photon 4-momentum is tracked
while numerically tracing the photon geodesic back in time from the
observer's grid to the surface of the disk (inner or outer). We
discard photons that are intercepted by the black hole or otherwise
miss the disk. Thus we properly accoint for the role of the black 
hole in absorbing photons emitted near the horizon. 
Here we assume that the transition region connecting
the two disks is optically thin (i.e. photons passing between the
two disk components near the transition radius don't "hit"
anything).  The validity of this assumption depends on the width of
the transition zone and upon how much material on average is present
within the transit region, neither of which is known at this time.

In this work we also ignore the illumination of the disk by line
photons emitted in other parts of the disk and the possible
reflection of those photons to the observer.  Because of
uncertainties in the emissivity function and the fact that the flux
in the line is a small fraction of the X-ray continuum in this
energy range, this is probably reasonable. Nevertheless, in future
work we plan to explore the effect of line reprocessing in
Bardeen-Petterson disks.

\subsection{Line Profiles}
\label{sec:profiles}

From the pixelated disk image, we generate a synthetic line profile
by making a histogram of the number of pixels in different frequency
bins. Each frequency bin is an accumulation of the contributions of
individual pixels and thus represents an integral over the disk. The
actual flux comes from weighting each bin by the emissivity
$\epsilon(r)$ and $g^4$; three factors of $g$ coming from the
relativistic invariant $I_\nu/\nu^3$, where $I_\nu$ is the
intensity, and the remaining factor arising because the line profile
is an integrated flux.  This weighting takes care of the effects of
Doppler motions and gravitational redshifts. The emissivity 
$\epsilon(r)$ is the {\em local} emissivity in the frame of the 
disk. Since some fraction of the photons emitted close to the black 
hole will be captured by it, $\epsilon(r)$ does not correspond 
directly to the energy received by the distant observer.

Throughout this paper we cast our results in terms of a 6.4 keV
rest-frame line energy. However, nothing in our methodology is tied
to this value. In this sense our results would be equally valid for
any other spectral line that might be emitted from the region of the
disk studied here. However, as will become clear, our results would
be considerably confused if two or more lines were emitted with
rest-frame energies separated by $\Delta E/E \lesssim 0.2$. This
could be particularly troublesome in accretion disks with a mixture
of H-like, He-like, and cold iron.

Throughout this work we assume an emissivity of the form
$\epsilon(r)\propto r^{-q}$ and generally use $q=2.5$. This is steeper 
than the $r^{-2}$ expected for a centrally illuminating source, yet 
shallower than the $r^{-3}$ expected from local energy dissipation 
within a disk. Obviously a
shallower radial dependence would increase the flux contribution of
the outer disk, whereas a steeper radial dependence would decrease
it. Regardless, provided a detectable amount of flux is emitted by
both disk components, our results will be qualitatively insensitive
to the choice of $q$ since the locations (in frequency or energy) of
the red and blue horns are insensitive to the specific flux
distribution \citep{cun75}. However, the strength of each feature,
and hence its detectability, {\em will} depend upon the value of
$q$. We explore this dependence further below.

Other details of our results may also depend on the form of the emissivity 
function we have chosen. For instance, in the case of a central compact 
source (or in so-called ``lamppost'' models), the change in the angle 
of incidence of the illuminating radiation 
beyond $r_{BP}$ would cause a change in the slope of 
$\epsilon(r)$. Accounting for such effects would require tracing the photon 
trajectories from their originating source, which is not done in the 
current work.

\begin{figure*}
%\plotone{Figure1.eps}
\plotone{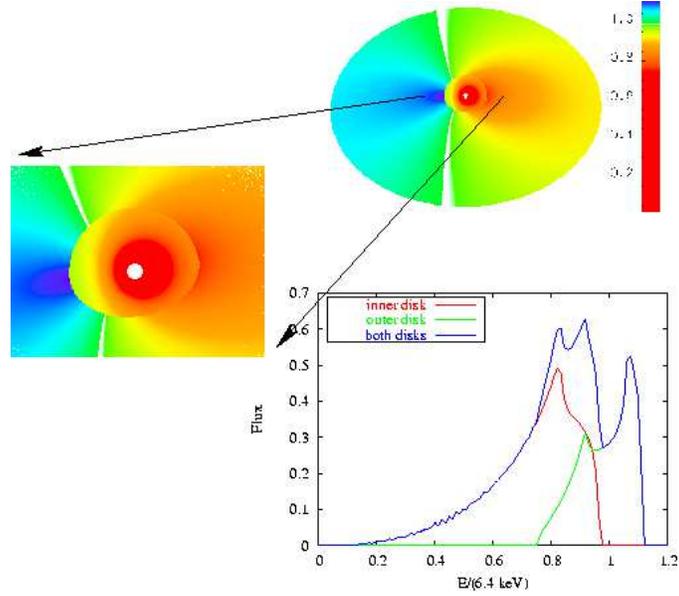}
\caption{Example disk image of a two-component Bardeen-Petterson
disk with $\beta=30^\circ$ and $r_{BP}=15r_g$. The observer is
located at $\phi_0=90^\circ$ (line-of sight perpendicular to the
line-of-nodes) and is inclined $45^\circ$ with respect to the outer
disk and $15^\circ$ with respect to the inner disk.}
\label{fig:disk1}
\end{figure*}

\begin{figure*}
%\plotone{lp2.eps}
\plotone{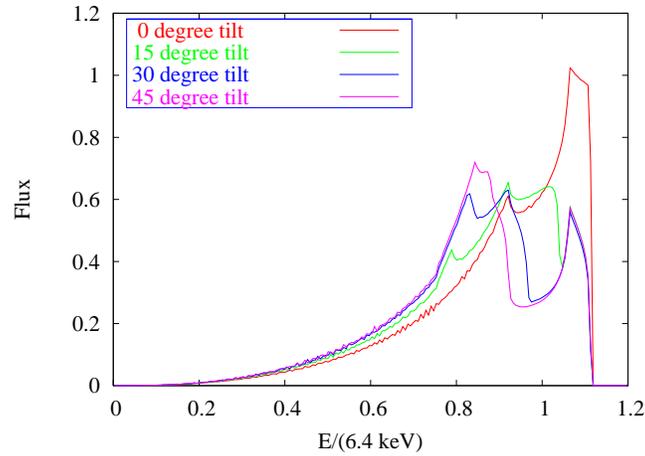}
\caption{Synthetic line profiles for a sample of two-component
Bardeen-Petterson disks with $r_{BP}=15r_g$ and various values for
the tilt $\beta$ in the range $0\le \beta \le 45^\circ$. The
observer is located at $\phi_0=90^\circ$ and maintains a constant
inclination of $i_{out}=45^\circ$ with respect to the outer disk.}
\label{fig:tilt}
\end{figure*}

Before considering specific black-hole systems, we first explore how
our synthetic line profiles depend on the various parameters of
interest. We start by testing the dependence of such profiles on the
tilt angle $\beta$ between the two disks. We consider four models
with values of $\beta$ in the range $0 \le \beta \le 45^\circ$. For
all of these models, the transition radius is set to $r_{BP}=15r_g$,
the inclination between the observer and the outer disk is fixed at
$i_{out}=45^\circ$, and the observer is located at $\phi_0=90^\circ$
(perpendicular to the line-of-nodes). Figure \ref{fig:disk1}
shows both the image of the disk and the line profile for the
$\beta=30^\circ$ case. For illustration, we have plotted the line 
profiles of each disk component separately as well as the composit 
line profile. This gives a good feel for how the remainder
of the line profiles are generated. Figure \ref{fig:tilt} shows the
composit line profiles for the four models with varying $\beta$. 
Clearly the $\beta \ne 0$
cases are easily distinguishable. The most
prominent signature of the Bardeen-Petterson profiles is the
presence of an "extra" horn, contributed by the inner disk
component. Since the observer remains fixed relative to the outer
disk in all these models, the blue edge $g_{max}$, which is
controlled here by the outer disk, remains fixed. However, the extra
blue wing from the inner disk is sensitive to the tilt $\beta$. In
these models, a larger $\beta$ translates into a smaller inclination
for the inner disk which, in turn, results in the corresponding blue
horn becoming redder. In fact, for the most tilted case considered,
the blue horn of the inner disk (located at $g\approx 0.84$) is
found at a lower energy than the red horn of the outer disk (located
at $g\approx 0.92$). The key point, though, is that the separation
of the various horns is a sensitive indicator of the tilt between
the two disk components. In other words, measuring this separation
provides a direct constraint on $\beta$, although, as we show below,
this conclusion does depend on the observer's location relative to
the line-of-nodes.

%\clearpage

%\clearpage
%\clearpage

\begin{figure*}
%\plotone{lp3.eps}
\plotone{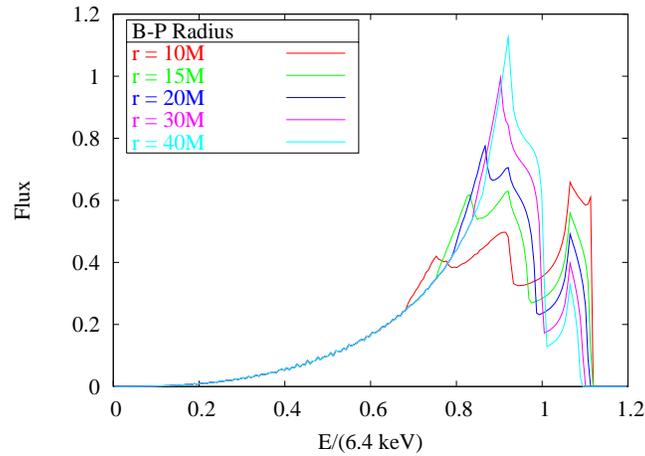}
\caption{Synthetic line profiles with the tilt fixed at
$\beta=30^\circ$ and the transition radius varied over the range
$10\le r_{BP} \le 40$. The observer is located at $\phi_0=90^\circ$
and is inclined $45^\circ$ with respect to the outer disk and
$15^\circ$ with respect to the inner disk.} \label{fig:radius}
\end{figure*}

Next, we explore the dependence of a set of synthetic
Bardeen-Petterson line profiles on the scale of the transition
radius $r_{BP}$. For these models, $\beta$ remains fixed at
$30^\circ$ and the observer is located at \{$\theta_0=15^\circ$,
$\phi_0=90^\circ$\}, giving inclinations of $15^\circ$ and
$45^\circ$ between the observer and the inner and outer disk
components, respectively. Figure \ref{fig:radius} shows the line
profiles for five values of $r_{BP}$ in the range $10 \le r_{BP}/r_g
\le 40$. For these models, it is not the blue horn that shows the
greatest variability, but rather the red horn. Thus, while the blue
horn is sensitive to the tilt $\beta$ between the disks, the red
horn is sensitive to the transition radius $r_{BP}$. This is because
the red horn closely tracks the inner radius of a given disk
component. Since increasing $r_{BP}$ is equivalent in our models to
increasing the inner radius of the outer disk component, varying the
transition radius causes a noticeable shift in the red horn of the
outer disk from $g\approx 0.76$ for $r_{BP}=10r_g$ to $g\approx
0.92$ for $r_{BP}=40r_g$. Note that for the set of parameters
considered here, the inner disk does not exhibit a red horn;
however, even if it did, it would not be very sensitive to a change
in the transition radius since the minimum radius of the inner disk
component would not change. Varying the transition radius also
changes the amount of flux coming from the inner and outer disk
components, so that the relative amplitudes of the various horns
provides another diagnostic of Bardeen-Petterson disks. A final
point to emphasize from Figure \ref{fig:radius} is that the
Bardeen-Petterson effect causes {\em detectable} changes in the iron
line profile even when $r_{BP} \ge 40 r_g$. This is an important
result, although as we show below, it is somewhat dependent on form
of the emissivity function.
%\clearpage
%\clearpage
%The line profile for a given Bardeen-Petterson disk will also depend
%on the specific location of the observer. Considering for the
%moment, only observers with a line-of-sight perpendicular to the
%line-of-nodes (i.e. $\phi_0=90^\circ$), we explore in Figure
%\ref{fig:latitude} the dependence of the line profiles on the
%inclination of the two disk components relative to the observer. For
%these models, the tilt between the disks and the transition radius
%remain fixed at $\beta=30^\circ$ and $r_{BP}=15r_g$, respectively.
%The observer is then moved in steps from $\theta_0=0$ to
%$\theta_0=90^\circ$. Here, since the tilt between the disks is
%fixed, the separation of the blue horns remains roughly constant,
%although the locations of these horns ($g_{max}$) shift. The
%locations of the red horns ($g_{min}$) remain nearly constant due to
%their weak dependence on inclination.

\begin{figure*}
\plotone{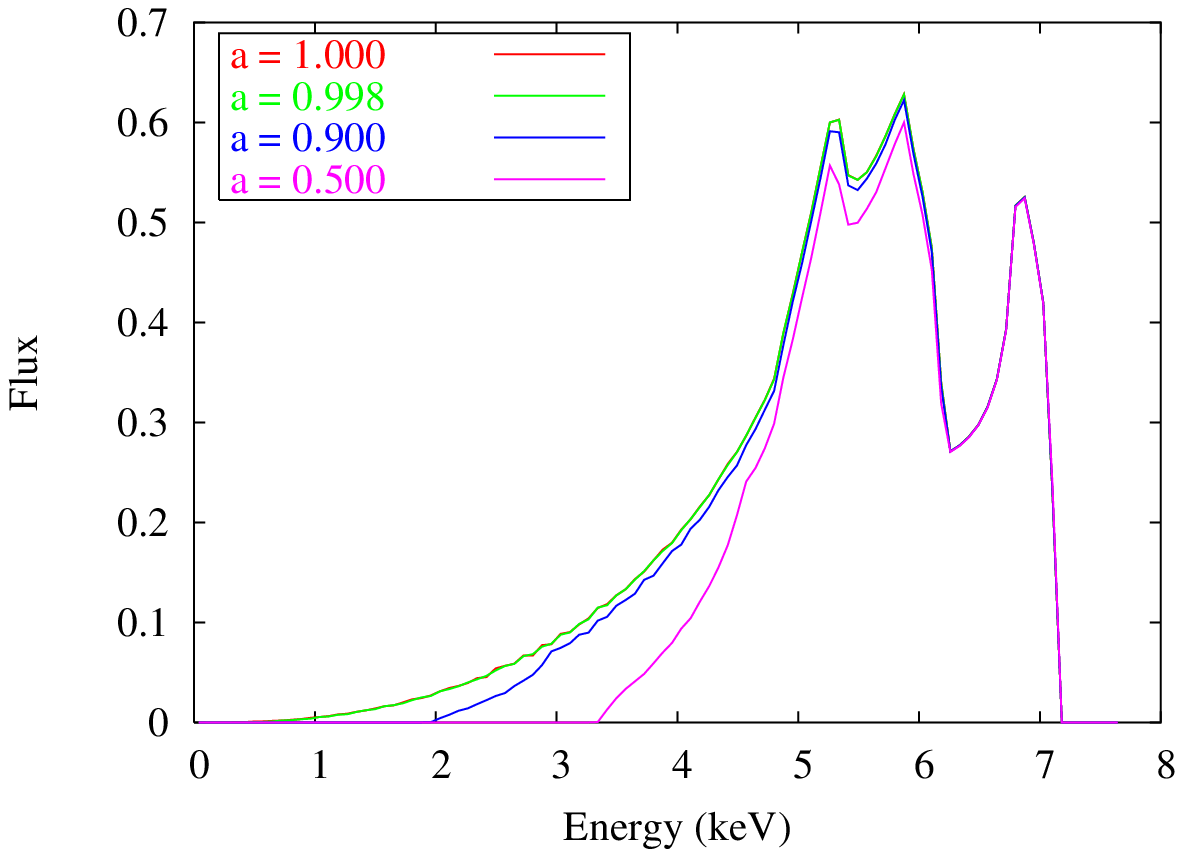}
\caption{Synthetic line profiles with the tilt,
transition radius, longitude of the observer, and inclination of the
observer relative to the outer disk fixed at $\beta=30^\circ$,
$r_{BP}=15r_g$, $\phi_0=90^\circ$, and $i_{out}=45^\circ$,
respectively. The emissivity assumes a form $\epsilon(r)\propto
r^{-q}$, with $q=2.5$. The spin of the black hole is varied over 
the range $0.5 \le a/M \le 1$.
} \label{fig:spin}
\end{figure*}

Next we vary the black hole spin parameter $a$. Although the 
Bardeen-Petterson transition radius is expected to be a function of $a$, 
in this work we treat them as independent parameters so that $r_{BP}$ 
remains fixed at $15r_g$ for the current set of models. The remaining 
parameters are also held at their canonical values ($\beta=30^\circ$, 
$i_{out}=45^\circ$, $\phi_0=90^\circ$, and $q=2.5$). Line profiles for 
four values of $a/M$ between 0.5 and 1 are shown in Figure 
\ref{fig:spin}. The biggest effect of changing $a$ is that the inner 
radius of the disk $r_{in}$ (assumed equal to $r_{ms}$) moves to 
larger radii for smaller $a$ ($r_{ms}=1$, 1.24, 2.32, and 4.23 for 
$a/M=1$, 0.998, 0.9, and 0.5, respectively). Since photons are not 
emitted from as deep in the potential well for larger $r_{in}$, the 
flux in the red tail is greatly diminished. Overall the flux 
contribution of the inner disk component is also reduced due to its 
decreased surface area.
%\clearpage
%\clearpage

\begin{figure*}
%\plottwo{lp4_1.eps}{lp4_2.eps}
\plotone{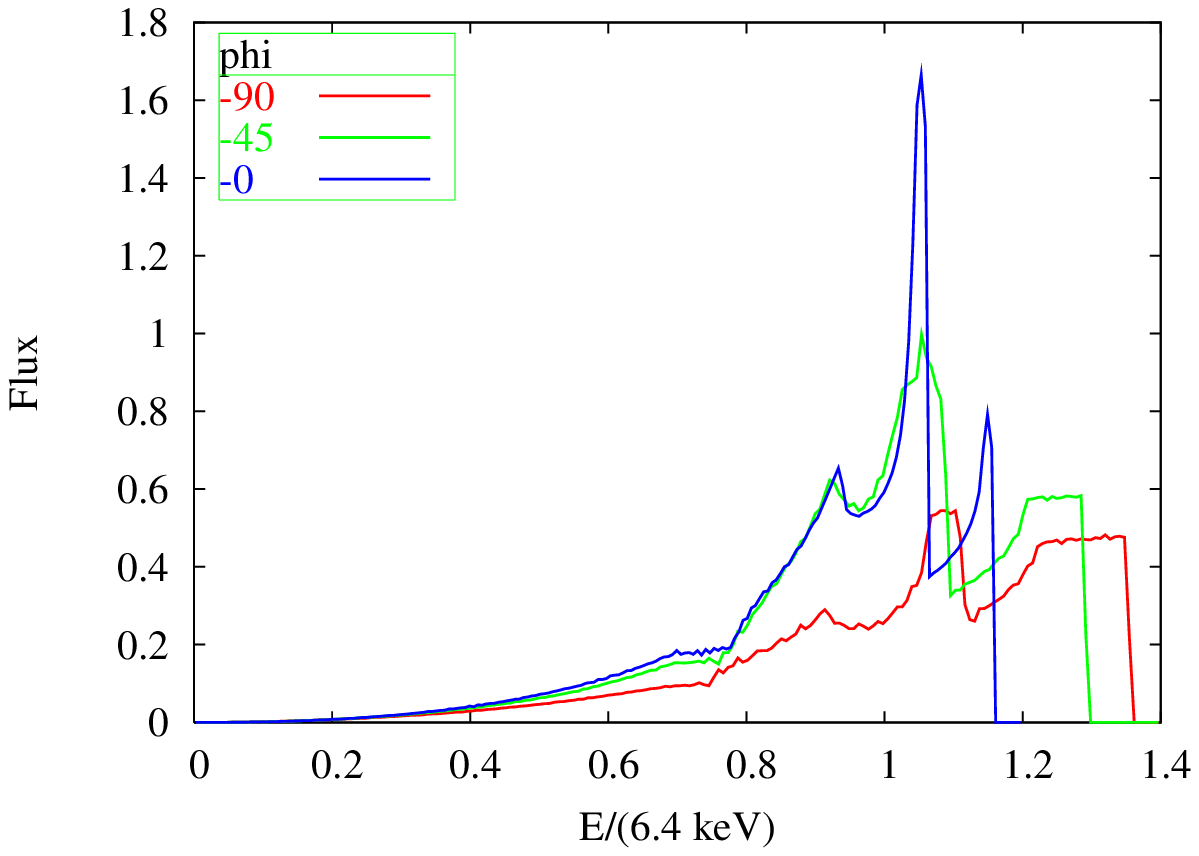}
\caption{Synthetic line profiles with the tilt,
transition radius, and inclination of the observer relative to the
outer disk fixed at $\beta=30^\circ$, $r_{BP}=15r_g$, and
$i_{out}=45^\circ$, respectively. The longitude of the observer is
varied over the range ({\em a}) $0\le \phi_0 \le 90$ and ({\em b})
$-90 \le \phi_0 \le 0$.} \label{fig:longitude}
\end{figure*}

\begin{figure*}
%\plotone{lp5.eps}
\plotone{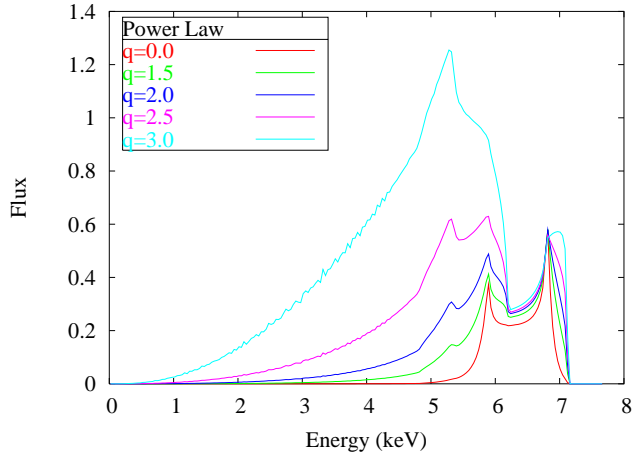}
\caption{Synthetic line profiles with the tilt,
transition radius, longitude of the observer, and inclination of the
observer relative to the outer disk fixed at $\beta=30^\circ$,
$r_{BP}=15r_g$, $\phi_0=90^\circ$, and $i_{out}=45^\circ$,
respectively. The emissivity assumes a form $\epsilon(r)\propto
r^{-q}$, with the numerical value of $q$ varied from 0 to 3. All of
these profiles are normalized using the peak flux in the blue horn
at $E=6.7$. 
The normalization factors are $2.73\times10^4$, 50, 6.73, 1, 
and 0.157 for $q=0$, 1.5, 2.0, 2.5, and 3.0, respectively.
} \label{fig:q}
\end{figure*}

So far we have only considered observers with a line-of-sight
perpendicular to the line-of-nodes ($\phi_0=90^\circ$).
However, the line profile for a given Bardeen-Petterson disk will
depend on the specific location of the observer, so we now consider
a set of line profiles made from the same Bardeen-Petterson disk
configuration, but seen from different viewing angles. To do this,
we maintain a constant inclination between the observer and the
outer disk while shifting the value of $\phi_0$. Effectively, the
observer is circling the angular momentum axis of the outer disk.
For these models, the tilt, transition radius, and inclination of
the outer disk relative to the observer remain fixed at
$\beta=30^\circ$, $r_{BP}=15r_g$, and $i_{out}=45^\circ$.  The
observer is shifted through $180^\circ$ from the positive $y$-axis
($\phi_0=90^\circ$) to the negative $y$-axis ($\phi_0=-90^\circ$).
Figure \ref{fig:longitude} shows the resulting line profiles. The
most noticeable change is the dramatic shift in the blue horn
attributed to the inner disk, which ranges from $g\approx 0.81$ for
$\phi_0=90^\circ$ to $g\approx 1.35$ for $\phi_0=-90^\circ$. The
contribution from the outer disk, on the other hand, remains roughly
constant since the observer maintains a constant orientation with
respect to this component and the transition radius remains fixed;
therefore, the red and blue horns attributed to the outer disk are
always located near $g\approx 0.92$ and $g\approx 1.05$,
respectively. Nevertheless, there are still small changes in the
outer disk contribution due mostly to shadowing by the inner disk.
%For reference, figure \ref{fig:disk2} shows the disk as it
%looks from \{$\theta_0=??^\circ$, $\phi_0=??^\circ$\}.
%\clearpage
\begin{figure*}
%\plottwo{lp4_1.eps}{lp4_2.eps}
\plotone{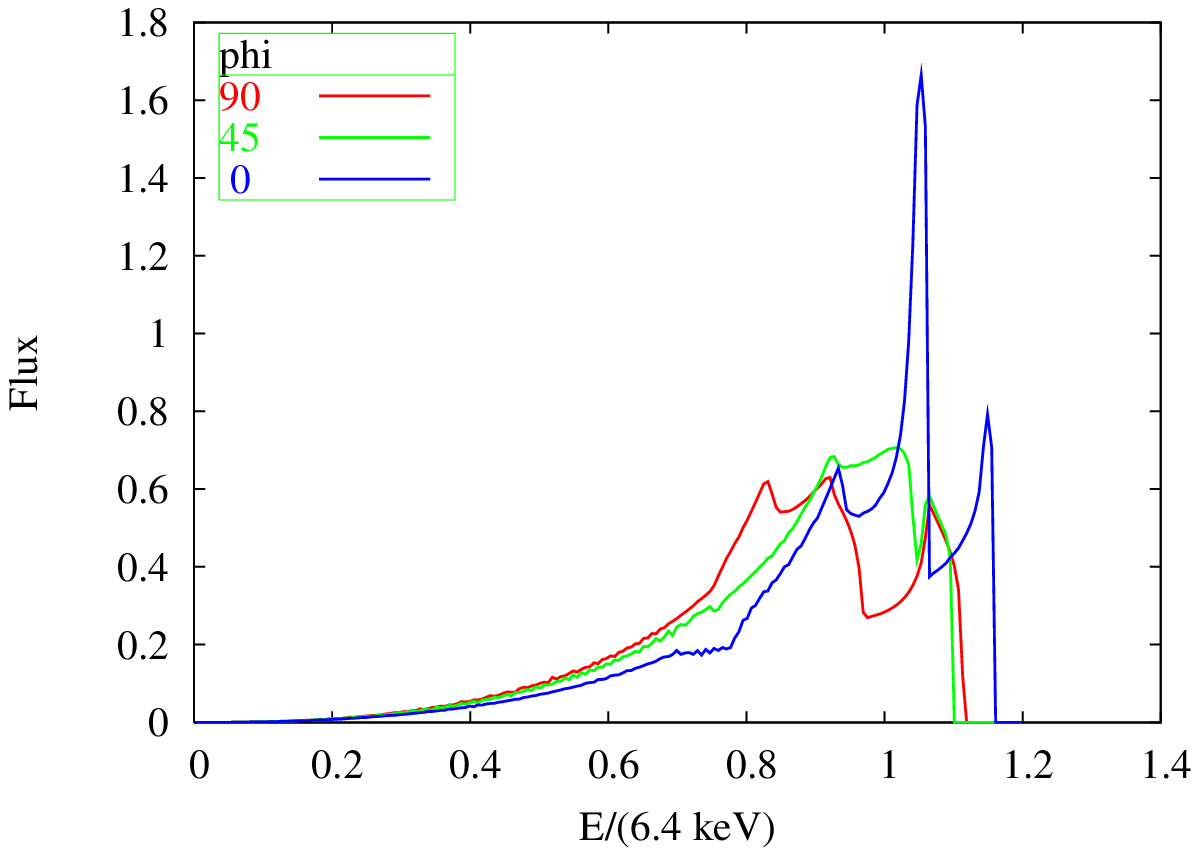}
\end{figure*}
%\clearpage
%\clearpage
%\clearpage
Thus far the results have shown that an iron line profile can be a 
sensitive probe for determining the relative tilt between two
components of a Bardeen-Petterson disk and the radius at which the
two components meet.  However, these results have all assumed an
emissivity of the form $\epsilon(r)\propto r^{-q}$ with $q=2.5$.
Although this is a reasonable choice based upon theoretical
expectations and is consistent with some fitted line profiles, it is
by no means the only plausible choice. Other values of $q$ are
certainly possible, as are other functional forms for $\epsilon$
(e.g. the "lamppost" model is a popular alternative). We finish this
section by considering a few other choices of $q$. Figure
\ref{fig:q} illustrates how the line profiles vary for choices of
$q$ from 0 to 3. In all models, $\beta=30^\circ$, $r_{BP}=15r_g$,
$i_{out}=45^\circ$, and $\phi_0=90^\circ$. For $q \le 2$, the
results are sensitive to our choice of $r_{out}$, which remains
fixed at $100r_g$ in all cases. The profiles are normalized by the
flux in the blue horn. It is clear from the figure that for $q
\lesssim 1$ the large outer disk component swamps the emission from
the inner disk. Only for steeper power-laws does the inner disk
contribution become apparent. If we were to continue this exercise
to consider even steeper power-laws, the inner disk component would
eventually swamp the outer disk contribution. It is in the
intermediate region, when both disk components are detectable, that
this technique is most useful as a probe of
Bardeen-Petterson disks. The exact range of $q$ for which this will
be true will depend on many parameters including $r_{min}$,
$r_{BP}$, $r_{max}$, and the inclination of the observer relative to
the two disk components. Nevertheless, we have shown that this
technique should work for a reasonable range of choices of all
parameters.
%\clearpage
%\clearpage

\section{Implications for Galactic Black Holes and Seyfert Galaxies}
\label{sec:examples}

In the previous section, we showed that iron-line profiles could, in
principle, be used as a diagnostic of Bardeen-Petterson disks.
Useful science could be extracted from such profiles in the form of
constraints on the relative tilt of Bardeen-Petterson disks and on
the location of the transition radius. However, the practical effect
of including the Bardeen-Petterson effect in modeling of iron-line
profiles is to add three new free parameters that must be fit: two
new parameters come from the tilt $\beta$ and transition radius
$r_{BP}$, themselves, and the third new parameter comes from
replacing a single inclination angle $i$ with the two angles
\{$\theta_0$, $\phi_0$\} necessary to describe the observer's
location relative to the two disk components. Given the many
uncertainties already inherent in fitting relativistically broadened
lines, this is not an entirely attractive prospect. Nevertheless, we
discuss some examples where there may be compelling reasons to make
the effort.

\subsection{Galactic Black Holes}
\label{sec:bhbs}

Often the inclination $i$ of an X-ray binary orbit can be determined
through careful observation and fitting of the photometric light
curve, the radial velocity curve, and other observables (e.g.
rotational velocity of the secondary), when the system is in X-ray
quiescence. In cases where the primary black hole is a microquasar,
the spin axis of the black hole can be inferred from the orientation
of the relativistic jet, which makes an angle $\theta_{jet}$ with
respect to the line-of-sight of the observer. If both $i$ and
$\theta_{jet}$ are reasonably constrained, then the position of the
observer ($\theta_0, \phi_0$) in our models is confined to the
family of points satisfying
\begin{eqnarray}
\theta_0 & = & \theta_{jet} \nonumber \\
\sin \phi_0 & = & \frac{\cos i - \cos \beta \cos \theta_{jet}}{\sin
\beta \sin \theta_{jet}} \label{eq:angles}
\end{eqnarray}
where we have utilized the law of cosines on a sphere.  In this way,
we can remove one of the extra degrees of freedom required to fit a
Bardeen-Petterson disk, leaving only the dependencies on $\beta$ and
$r_{BP}$. At this time, this approach probably gives the best hope
for constraining the Bardeen-Petterson transition radius. (Spectral
modeling of the continuum may also provide important constraints.)
Following this procedure, we discuss two GBHs that are likely
candidates for Bardeen-Petterson disks and provide sample synthetic
line profiles of each. For now these synthetic line profiles are
illustrative in nature since we do not fit actual observations.
However, we do discuss the qualitative similarities between our
synthetic lines and actual line profiles in the literature
\citep{bal01,mil04}.

\subsubsection{GRO J1655-40}
\begin{figure*}
%\plotone{plot6c.epsf}
\includegraphics[width=4in, angle=-90]{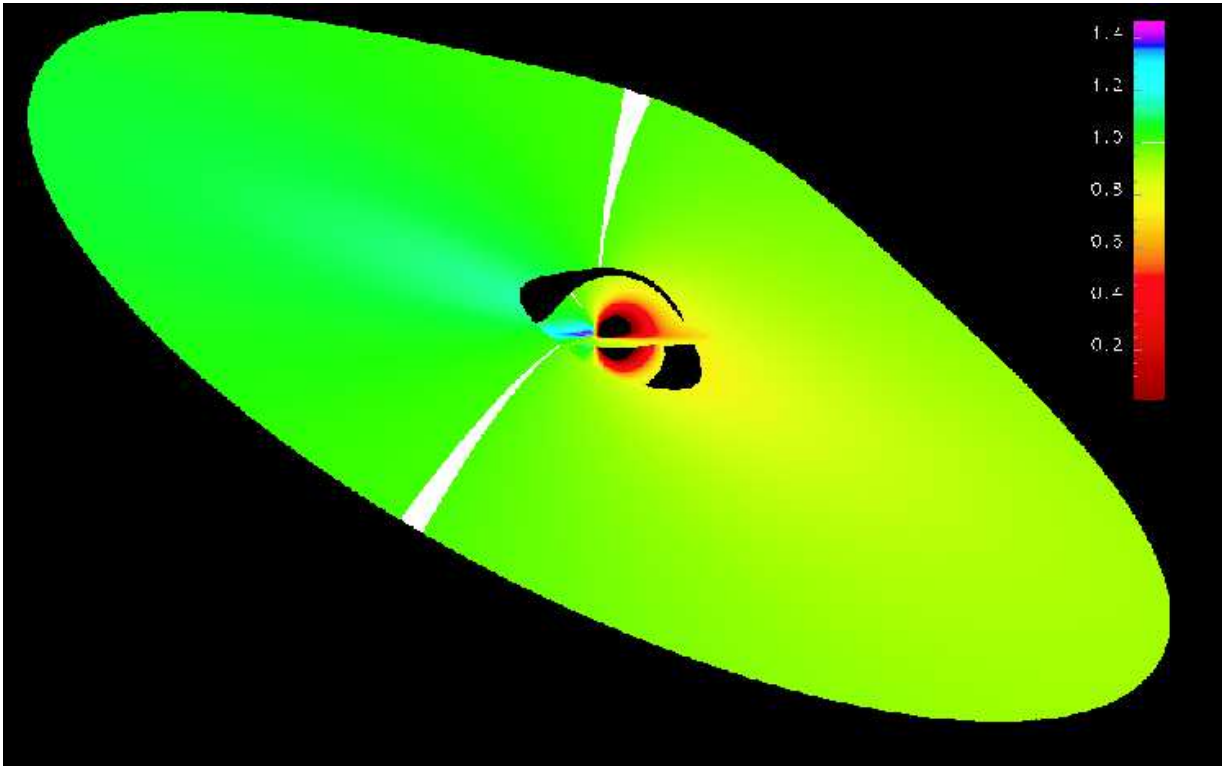}
\caption{Disk image of GRO J1655-40 for an observer at \{$\theta_0$,
$\phi_0$\}=\{$85^\circ$, $32^\circ$\} ($\beta=30^\circ$). Using
observational constraints we fix the inclinations of the inner and
outer disk components at $i_{in}=85^\circ$ and $i_{out}=70.2^\circ$,
respectively. The transition radius is set at $r_{BP}=15r_g$.}
\label{fig:GRO_disk}
\end{figure*}

\begin{figure*}
%\plottwo{lp_GRO81co.eps}{lp_GRO32co.eps}
\plotone{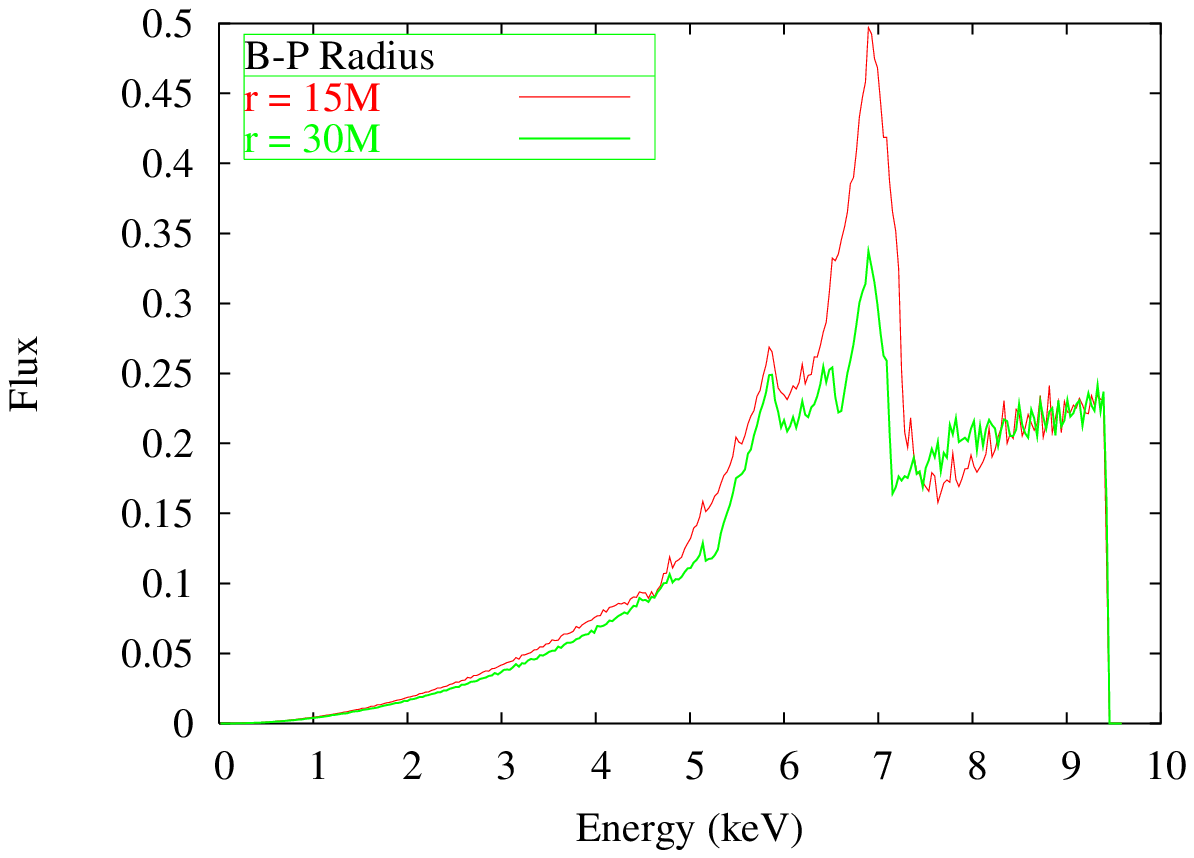}
\caption{({\em a}) Synthetic line profiles for GRO J1655-40 for an
observer at \{$\theta_0$, $\phi_0$\}=\{$85^\circ$, $81^\circ$\}
($\beta=15^\circ$) and ({\em b}) \{$\theta_0$,
$\phi_0$\}=\{$85^\circ$, $32^\circ$\} ($\beta=30^\circ$). In all
cases, we assume the black hole is maximally rotating,
$r_{max}=100M$, and $\epsilon \propto r^{-2.5}$. In each case, we
present profiles for two values of the transition radius
($r_{BP}=\{15,30\}r_g$).} \label{fig:GRO_lines}
\end{figure*}

GRO J1655-40 is one of the best studied and most tightly constrained
GBHs, with a binary period of $2.62191\pm0.00020$ days, a primary
(black-hole) mass of $6.3\pm0.5 M_\odot$, and a binary inclination
of $i=70.2\pm1.9^\circ$ \citep{gre01}. It has also been previously
identified as a likely candidate for a Bardeen-Petterson disk
\citep{fra01a}. This characterization comes from noting the
discrepancy between the binary inclination and the orientation of
the highly collimated bipolar relativistic jet seen in this source
during a 1994 outburst
\citep[$\theta_{jet}=85\pm2^\circ$;][]{hje95}. If we assume that the
orientation of the jet is fixed by the angular momentum axis of the
black hole, then the tilt between the binary orbital plane and the
black hole symmetry plane is $\beta \ge 10.9^\circ$. Due to the
remaining degree of freedom to select the precise location of the
observer (i.e. choosing $\phi_0$ in our models), we can not place a
tight upper limit on the tilt. The formal upper limit is $\beta \le
(180-10.9)^\circ=169.1^\circ$. Nevertheless, we argue on physical
grounds that $\beta \le 30^\circ$ is probably a reasonable upper
limit. For instance, values of $\beta>90^\circ$ would imply that the
inner disk is orbiting in the opposite sense of the outer disk.
Although such large tilts may be possible in some systems (e.g. 
systems that have undergone binary capture or binary replacement), such
counter-rotating disks seem a highly unlikely scenario. In fact, it
is hard to imagine that a Bardeen-Petterson disk structure could be
supported if $\beta$ were much larger than about $45^\circ$, as this
would require a dramatic reorientation of the angular momentum of
the gas in the transition region. Adopting a limit of $\beta \le
30^\circ$ and using equation \ref{eq:angles} to fix the observer's
location \{$\theta_0$, $\phi_0$\}, we consider two possible tilt
angles for GRO J1655-40: $\beta=15^\circ$ and $30^\circ$. The
corresponding observer locations are \{$\theta_0$,
$\phi_0$\}=\{$85^\circ$, $81^\circ$\} and \{$85^\circ$,
$32^\circ$\}. In both cases, the inclinations of the inner and outer
disks are $i_{in}=85^\circ$ and $i_{out}=70.2^\circ$, respectively,
consistent with the observations. Thus the inner disk is viewed more
edge on than the outer disk, although both are viewed from fairly
low latitudes, as shown in the disk image in Figure
\ref{fig:GRO_disk}. Notice that the inner disk is viewed from such a
low latitude that photons from the bottom of the disk are bent
enough to propagate through the gap at the transition radius and
make it to the observer. Although this is an interesting
illustration of light-bending effects, it also highlights the
limitations of our assumptions of an infinitesimally thin disk and
an optically thin transition radius. In a real system, these photons
would likely be blocked from view by the flaring of the disk or
optically thick gas in the transition region. Nevertheless, we
include these photons in the synthetic line profiles shown in Figure
\ref{fig:GRO_lines}. The four lines in the figure correspond to the
two observer locations and two values for the transition radius:
$r_{BP}=\{15,30\}r_g$. All of the models assume a maximally rotating
black hole, an outer radius of $r_{max}=100M$, an observer at
$r=500M$, and a disk emissivity of the form $\epsilon \propto
r^{-2.5}$. The noise in the profiles is a result of the poor
resolution of the inner disk, which again is seen nearly edge-on.
%\clearpage
\clearpage
\begin{figure*}
%\plottwo{lp_GRO81co.eps}{lp_GRO32co.eps}
\plotone{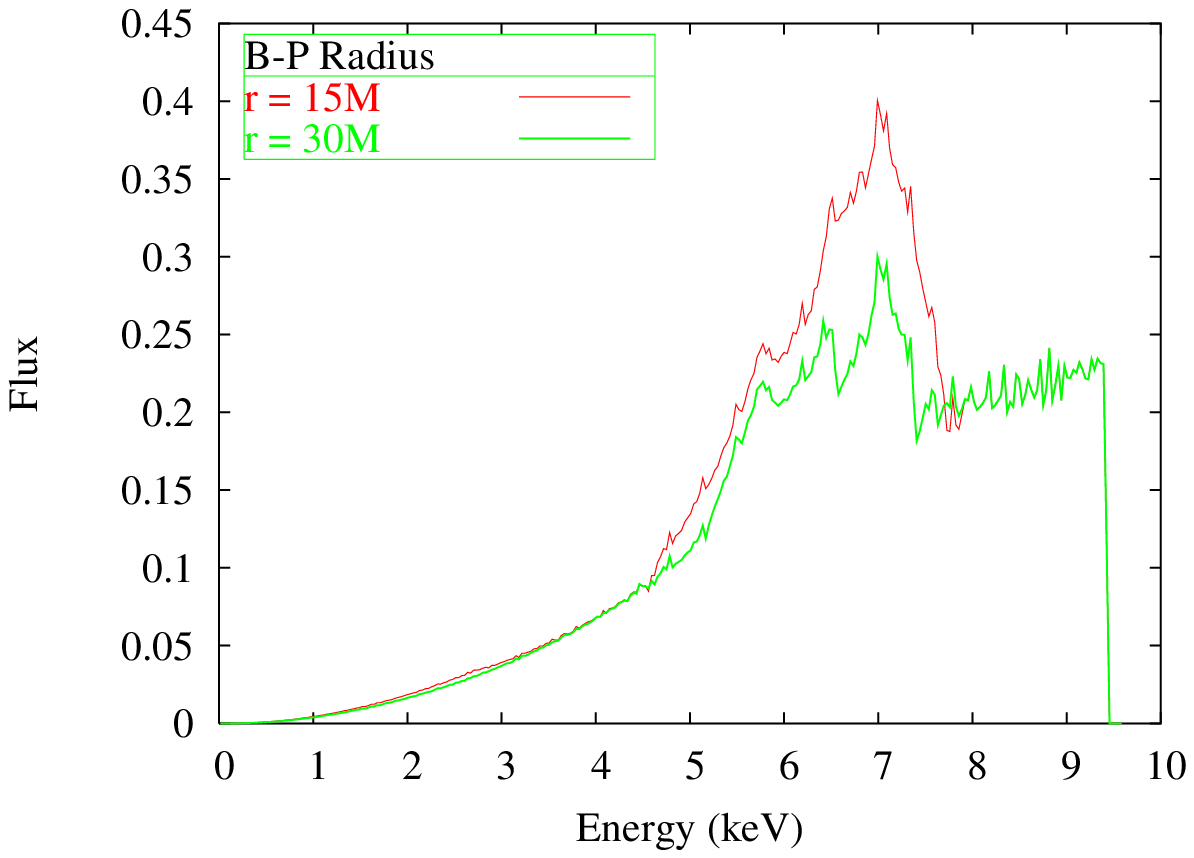}
\end{figure*}
%\clearpage
%\clearpage
The peak in the synthetic line profile at $E\approx7$ keV is
attributable to the blue horn of the outer accretion disk component
overlaid on the much broader line from the inner disk. The energy of
this peak is very close to the peak line energy seen in the {\em
ASCA} observations of this source \citep{mil04}. Assuming a
rest-frame line energy of 6.4 keV, this observed line energy
requires a disk inclination of $i\gtrsim45^\circ$, consistent with
all other indications that this system is seen at high inclination.
Rest-frame line energies of 6.7 or 6.9 keV (from highly ionized
iron) could be fit with smaller inclinations, but would require the
emitting region to be far from the hole.

The extremely high inclination of our model inner disk leads to a
very large value for the blue maximum ($g_{max}=1.5$ or $E_{max}=
9.4$ keV for a 6.4 keV line), below which there is a shallow blue
plateau. Although such high-energy line emission is not apparent in
the observed spectrum of GRO J1655-40 \citep{mil04}, this does not
necessarily rule out a highly inclined inner disk component. Such a
component could be present, yet have its emission obscured through
self-absorption by the finite-thickness of the disk itself (unlike
the infinitesimally thin disks considered here). Or it could be that
the high-energy line emission is hidden by absorption features. This
is a potential problem for all relativistically broadened lines, not
just those from Bardeen-Petterson disks. Absorption is particularly
problematic at energies above about 7 keV.

\subsubsection{XTE J1550-564}
\begin{figure*}
%\plottwo{plot7a.epsf}
\includegraphics[width=4in, angle=-90]{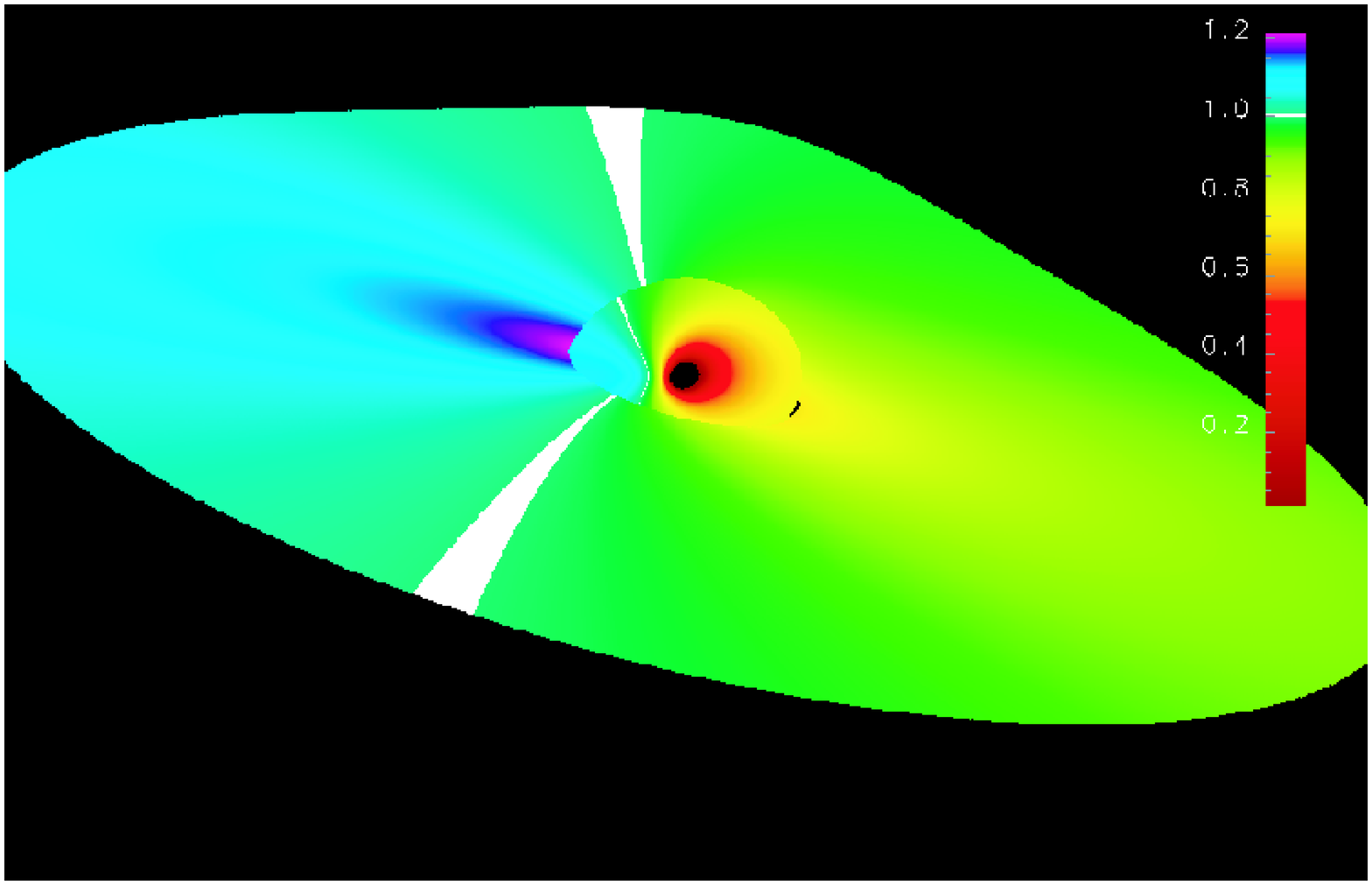}
\caption{Disk image of XTE J1550-564. We use observational
constraints to fix the inclinations of the outer disk component at
$i_{out}=72.2^\circ$. We assume the inner disk is inclined
$i_{in}=45^\circ$, consistent with the observational constraint
$i_{in}<53^\circ$. The transition radius is fixed at $r_{BP}=15r_g$
and the observer is located at \{$\theta_0$,
$\phi_0$\}=\{$45^\circ$, $-60^\circ$\}.} \label{fig:XTE_disk}
\end{figure*}

\begin{figure*}
%\plotone{lpXTEco.eps}
\plotone{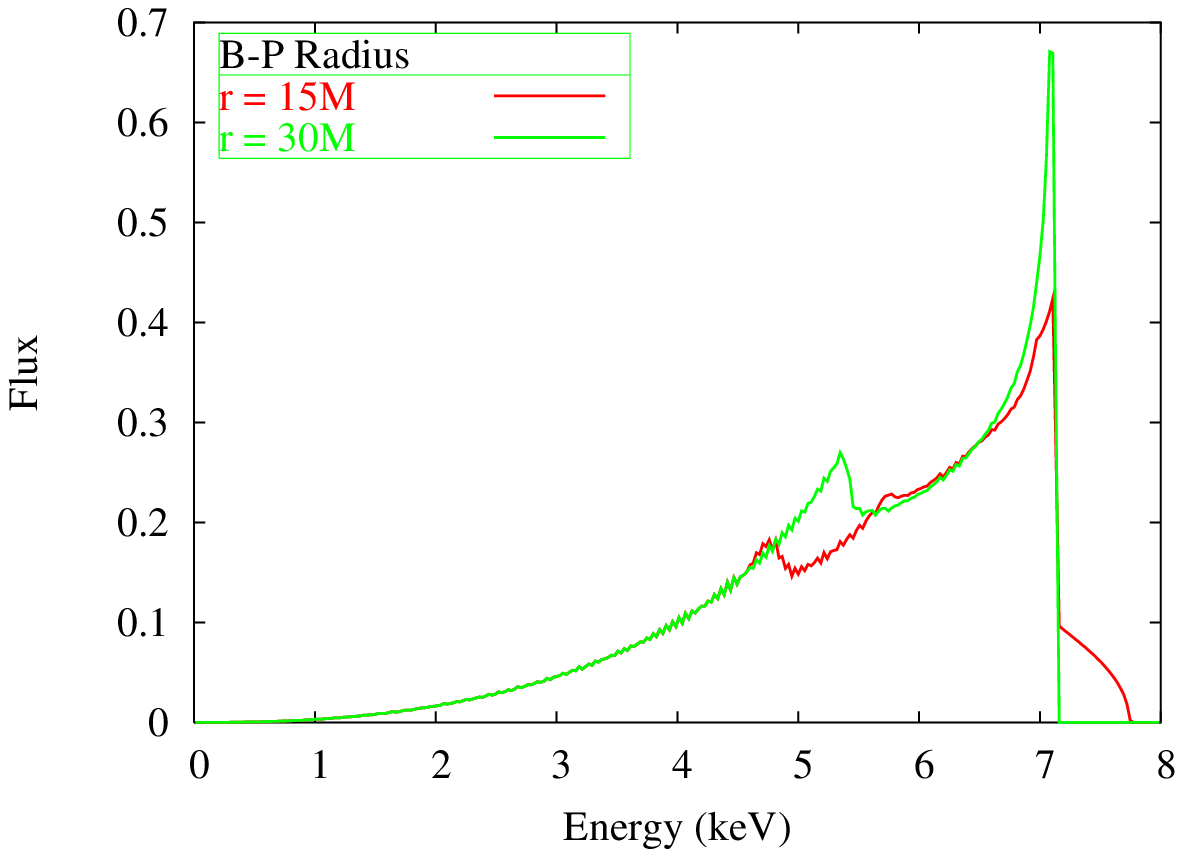}
\caption{Synthetic line profiles for XTE J1550-564. For these, we
assume the black hole is maximally rotating, $r_{max}=100M$,
$\epsilon \propto r^{-2.5}$, and $\beta=30^\circ$. We present
profiles for two values of the transition radius
($r_{BP}=\{15,30\}r_g$).} \label{fig:XTE_line}
\end{figure*}

XTE J1550-564 has many similarities to GRO J1655-40, although its
parameters are generally not as well constrained.  Despite the
larger uncertainties, it is still clear that the system hosts a
black-hole primary ($8.36M_\odot \le M_1 \le 10.76 M_\odot$).  The
binary period ($1.552\pm0.010$ days) and inclination ($67.0^\circ
\le i \le 77.4^\circ$) are also reasonably well known. Similar to
GRO J1655-40, XTE J1550-564 is a microquasar.  During a 1998
outburst, large scale relativistic jets were observed in both X-ray
and radio wavelengths \citep{han01,cor02}. The radio components
displayed apparent separation velocities of $>2c$ \citep{han01}.
Although the exact orientation of the jets was not confirmed, the
apparent superluminal motion itself constrains the angle of the jet
relative to the line-of-sight to be $\theta_{jet} < 53^\circ$.
Similar to GRO J1655-40 and V 4641 Sgr, this jet orientation is in
disagreement with the inclination of the binary orbit; in this case,
the implied tilt is $\beta \ge 14^\circ$. Although the limit on this
tilt is very similar to the limit for GRO J1655-40 and the binary
inclinations are quite similar, there is a key difference between
the two systems: for GRO J1655-40, the implied inner disk is viewed
more edge-on than the outer disk, whereas for XTE J1550-564, the
orientation is reversed and the outer disk is viewed more edge-on.
Since the superluminal motion of the jet only gives us an upper
limit on $\theta$, we are unable to use equation \ref{eq:angles} to
further constrain the observer location for XTE J1550-564.
Nevertheless, for illustration, in Figure \ref{fig:XTE_disk} we
assume a jet angle $\theta_{jet}=45^\circ$ and a tilt
$\beta=30^\circ$. The corresponding observer locations is
\{$\theta_0$, $\phi_0$\}=\{$45^\circ$, $-60^\circ$\}. The
inclinations of the inner and outer disks are $i_{in}=45^\circ$ and
$i_{out}=72.2^\circ$, respectively. Again we assume a maximally
rotating black hole, an outer radius of $r_{max}=100M$, an observer
at $r=500M$, and a disk emissivity of the form $\epsilon \propto
r^{-2.5}$. We consider two values of $r_{BP}=\{15,30\}r_g$. Since
the outer disk component for XTE J1550-564 is nearly identical to
what we used for GRO J1655-40, it is not surprising that its
contribution to the line profiles is quite similar (compare Figures
\ref{fig:XTE_line} and \ref{fig:GRO_lines}). The differences between
the two figures are mainly attributable to the inner disk component
and to differences in the shadowing of the various models.
%\clearpage
%\clearpage
%\clearpage
The synthetic spectrum of XTE J1550-564 in Figure \ref{fig:XTE_line}
has a peak at $E\approx7$ keV. Some of the photons in this peak come
from the outer edge of the inner disk component ($i_{in}=45^\circ$),
while the rest come from the inner edge of the outer disk
($i_{out}=72.2^\circ$). This value is roughly consistent with the
peak of the observed line in XTE J1550-564 at $E\approx 6.8$ keV
\citep{mil04}. For a rest frame energy of 6.4 keV, this requires a
disk inclination $i\gtrsim 45^\circ$. For $r_{BP}=15r_g$, the
synthetic spectrum also shows some excess emission in the range
$E=7-8$ keV. This comes from the inner edge of the outer disk
component and is not seen for $r_{BP} \gtrsim 30r_g$ since the
Keplerian velocity of the orbiting gas is not high enough in such a
case. Interestingly, excess emission above the continuum fit also
appears over the same energy range in the observations of XTE
J1550-564 \citep{mil04}. Line flux in this energy range requires
a disk inclination of $i\gtrsim 60^\circ$, consistent with the
binary inclination and the inclination of the outer disk component
in our models. Although we have not formally fit the data for XTE
J1550-564, it is clear that the observed profile is consistent with
a Bardeen-Petterson disk configuration.

\subsection{Seyfert Galaxies}
Although GBHs may provide the best constrained systems for studying
iron-line profiles, at this time their data do not have as high of a
signal-to-noise value as the observations of Seyfert 1 galaxies. The high
quality of Seyfert 1 data makes it appropriate to at least consider
this class of object as part of this study. Furthermore, there is a
potential puzzle that a Bardeen-Petterson disk might help rectify:
According to the unified model of active galactic nuclei, type 1
Seyfert galaxies are believed to be viewed at low inclination angles
($i\lesssim30^\circ$) in order that they not be obscured by the dust
torus presumed to surround the accretion disks at large radii.
However, such low inclinations would prohibit the blue wing of an
iron line from extending much above its rest-frame energy ($g_{max}
\lesssim 1$). Therefore, the presence of detectable emission above
6.4 keV is a potential problem. It has usually been interpreted as evidence
that the emission is coming from highly ionized states of iron (Fe
XXIV-XXVI). However, we consider a 
Bardeen-Petterson profile such as the ones shown in this work 
to be a potentially more
conservative, alternative explanation. In the context of the unified model,
the outer disk of a Seyfert 1 galaxy would be required to have a low 
inclination, but the Bardeen-Petterson model would allow the inner disk 
to take on a broader range of
inclinations. In particular, a higher inclination for the inner disk
would allow the blue wing to extend above 6.4 keV without resorting
to higher ionization states.

\subsubsection{MCG--6-30-15}
One of the best studied iron-line sources is the nearby type 1
Seyfert galaxy MCG--6-30-15. It was the first source to show a
relativistically broadened iron-line feature \citep{tan95} and
remains one of the most cited examples. In fact, the data for this
source are of such high quality, they actually reveal a complex set
of features that can not be adequately fit as a single line
\citep{fab02,bal03}. The key features include a strong, broad peak
near 6.4 keV, a very extended red tail down to $E\lesssim 3$ keV,
and an ``extra'' horn or peak near 6.9 keV. As mentioned above, a
6.9 keV emission feature is a potential problem for Seyfert galaxies in the
context of the unified model since disks seen at low inclinations
(like Seyfert-1s) can not produce strongly blue-shifted lines. Thus
this feature is often attributed to highly ionized states of iron
(Fe XXIV-XXVI). Alternatively we show that these complex features can
be qualitatively explained using a ``single'' 6.4 keV iron line from a
Bardeen-Petterson disk. The key principle is that we can satisfy the
unified model of AGN provided we model the outer disk component with
a small inclination while we are free to choose a broader range of 
inclinations for the inner disk. We suggest therefore that, in the
case of MCG--6-30-15, it is the outer component that produces the
core line near 6.4 keV while an inner disk component produces both the
extended red wing and the 6.9 keV feature. In order to do so from a 6.4 keV
rest-frame line, the inner disk must be inclined at least $45^\circ$
with respect to the observer \citep{par01}. If we assume that the
outer disk is inclined $30^\circ$, this gives a tilt $\beta \ge
15^\circ$ depending on the angle between the line-of-sight and
line-of-nodes ($\beta=15^\circ$ if $\phi_0=90^\circ$; $\beta>15^\circ$ if 
$\phi \ne 90^\circ$).

\begin{figure*}
%\plotone{lp_MCG.eps}
\plotone{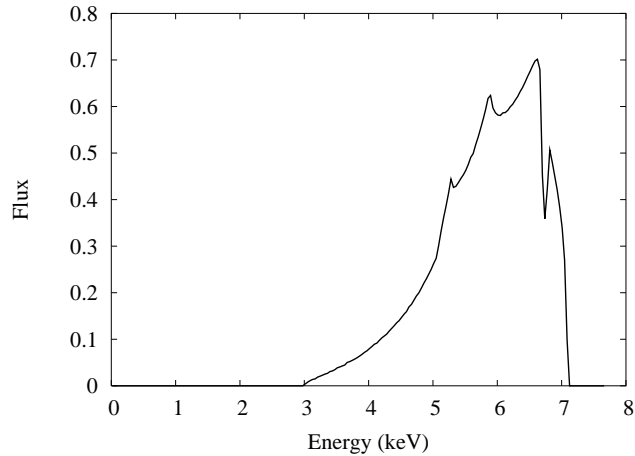}
\caption{Synthetic line profile providing a qualitative fit to
MCG--6-30-15. For this model, the disk parameters are
$r_{min}=4.2M$, $r_{BP}=20M$, $\beta=15^\circ$, $r_{max}=100M$, and
$\epsilon \propto r^{-2.5}$. The observer is located at
$\phi_0=90^\circ$ and has inclinations of $i_{out}=30^\circ$ and
$i_{in}=45^\circ$ with respect to the outer and inner disk
components, respectively.} \label{fig:MCG}
\end{figure*}

In Figure \ref{fig:MCG} we show a synthetic line profile that
qualitatively matches the data for MCG--6-30-15 \citep{fab02}. This
profile assumes the observer's line-of-sight is perpendicular to the
line-of-nodes ($\phi_0=90^\circ$) so that the tilt is
$\beta=15^\circ$. Unlike most of our previous models, in which the black hole
was maximally rotating, here we assume a spin $a/M=0.5$, giving an
inner disk radius of $r_{min}=4.2M$.  The transition radius is taken
to be $r_{BP}=20M$. As in previous models, we assume an emissivity
of the form $\epsilon \propto r^{-2.5}$. We note, however, that 
this emissivity may 
not be steep enough to reproduce the observed red wing flux. 
Nevertheless, it is worth point out that our relatively simple 
physical model is successful in matching 
many of the unusual features of the MCG--6-30-15 line profile.
%\clearpage
%\clearpage
As described earlier, a nice property of our model is that it
provides a natural explanation for why the flux in the line core of
MCG--6-30-15 does not match the flux in the red wing: these two flux
contributions come from different disk components with different
inclinations relative to the observer (in this case, the outer disk
is seen more face on).

\section{Conclusions}
\label{sec:conclusions} We set out in this work to demonstrate that
relativistically broadened iron lines can reveal important details
about Bardeen-Petterson disks in GBHs and AGN. In support of this
goal, we showed that such spectral lines are sensitive to the two
parameters of greatest interest in Bardeen-Petterson disks: the tilt
between the two disk components and the transition radius where they
meet. At this time, this approach probably gives the best hope for
determining these parameters, although spectral modeling of the
continuum can also provide some constraints.

We applied our results to a small set of potentially interesting
systems. One class of such systems is microquasars. These objects
are often better constrained in terms of the relevant parameters
than other potential targets such as AGN. Any fits to iron-line
profiles in microquasars must be consistent with observations of the
binary orbital parameters, the orientation of the relativistic jet,
and spectral fitting of the thermal (accretion disk) component. This
kind of consistency check will be particularly helpful in ruling out
competing effects. Unfortunately, current observations of iron lines
from microquasars do not have sufficiently high signal-to-noise to
resolve the extra features that are characteristic of
Bardeen-Petterson disks. Nevertheless, current or future missions
should be able to achieve the required sensitivity. In anticipation
of this, we have provided predicted spectra of two likely
Bardeen-Petterson microquasar candidates: GRO J1655-40 and XTE
J1550-564.

Although AGN are generally not as well constrained as microquasars,
they nevertheless represent another important class of objects in
the context of this work. Furthermore, they have historically
yielded higher quality line profiles. In this work we focused on one
example: MCG--6-30-15. Here we showed that a Bardeen-Petterson disk
profile could qualitatively explain the complicated line profile
observed in that system, {\em without resorting to extra lines from
other ionization states of iron}. Although other models have been
proposed that produce similar features, we feel the
Bardeen-Petterson explanation is an intriguing possibility. The
discovery of a Bardeen-Petterson disk in an AGN would reveal a
wealth of information about the environment of the accretion flow
and the history of the system.

The most restrictive limitation of this technique is that, 
to be fully exploited, both
components of the Bardeen-Petterson disk must produce detectable
levels of line flux. Provided this condition is met, we have shown
that the line profiles exhibit unique signatures, such as inverted
(red-horn-dominant) lines or lines with more than two horns, that
are not easily mimicked by other effects. Nevertheless, there are
competing effects one must be aware of. Probably the most important
are line blending, for instance from low and high ionization states
of iron, and disk warping due to mechanisms other than
Lense-Thirring precession. Line blending will be particularly
confusing if two or more lines are emitted with rest-frame energies
separated by $\Delta E/E \lesssim 0.2$. In such cases, one will have to rely
on other observational and physical constraints of the system to
disentangle the possibilities.

In closing, we would like to suggest that it might be fruitful in
the future to conduct a systematic search for Bardeen-Petterson
features among all observations of relativistically broadened iron
lines. Such a search could yield important statistical bounds on
questions such as: How many systems have tilted accretion disks?
What is the average tilt in such systems? What is the average
transition radius? Answering these questions will give important
clues toward understanding accretion-disk physics and the
evolutionary history of accreting black holes.

\begin{acknowledgements}
P.C.F. gratefully acknowledges useful discussion and feedback on
this work from S. Davis and O. Blaes. We thank the anonymous 
referee for suggesting important improvements to the manuscript. 
Funding support for P.C.F.
was provided by NSF grant AST 0307657. 
W.A.M. wishes to acknowledge support
from FAU's Office of Graduate Studies and from an LDRD/ER grant from LANL.
\end{acknowledgements}

%\clearpage
\bibliographystyle{apj}
\bibliography{myrefs}

%\clearpage

%%%%%%%%%%%%%%%%%%%%%% FIGURES %%%%%%%%%%%%%%%%%

%%%%%%%%%%%%%%%%%%%% TABLES %%%%%%%%%%%%%%%%%%%%%

\end{document}